\documentclass[11pt,preprint,graphicx]{aastex}
\def\etal{\it et al. \rm }

\begin{document}

\title{The Age of Ellipticals and the Color-Magnitude Relation}

\author{James Schombert}
\affil{Department of Physics, University of Oregon, Eugene, OR 97403;
js@abyss.uoregon.edu}

\author{Karl Rakos}
\affil{Institute for Astronomy, University of Vienna, A-1180, Wien, Austria;
karl.rakos@chello.at}

\begin{abstract}

Using new narrowband color observations of early-type galaxies in
clusters, we reconstruct the color-magnitude relation (CMR) with a higher
degree of accuracy than previous work.  We then use the spectroscopically
determined ages and metallicities from three samples (Trager \etal 2008,
Thomas \etal 2005, Gallazzi \etal 2006), combined with multi-metallicity
SED models, to compare predicted colors for galaxies with young ages (less
than 8 Gyr) with the known CMR.  We find that the CMR cannot by reproduced
by the spectroscopically determined ages and metallicities in any of the
samples despite the high internal accuracies to the spectroscopic indices.
In contrast, using only the $<$Fe$>$ index to determine [Fe/H], and assuming
a mean age of 12 Gyr for a galaxy's stellar population, we derive colors
that exactly match not only the color zeropoint of the CMR but also its
slope.  We consider the source of young age estimates, the H$\beta$ index,
and examine the conflict between red continuum colors and large H$\beta$
values in galaxy spectra.  We conclude that our current understanding of
stellar populations is insufficient to correctly interpret H$\beta$ values
and that the sum of our galaxy observations supports an old and monolithic
scenario of galaxy formation.  {\it This result has a devastating impact on
every study that has used the H$\beta$ index to calculate galaxy age, as
the use of the H$\beta$ versus MgFe diagram will result in incorrectly
deduced young ages}.

\end{abstract}


\section{INTRODUCTION}

The fundamental concepts in the fields of cosmology and galaxy formation
hinge on the one defining characteristic of galaxies, their age.  However,
a galaxy's age is a difficult parameter to define observationally.  For
example, in the current cold dark matter paradigm ($\Lambda$CDM, Bahcall
\etal 1999), a galaxy's age is the point in time when the non-baryonic
matter becomes gravitational bound.  These dark matter halos, then, induce
the infall of baryonic material, that forms the luminous galaxies that we
observe.  Typically, a collection of raw dark and baryonic matter is
impossible to date as they contain no signatures of age (i.e. an observable
clock).  Only if, and when, the baryonic matter collapses and initiates
star formation do we have any measure of the amount of time since the
formation epoch.

A complication arises if the time of initial star formation does not
immediate follow the epoch of gravitational collapse.  If star formation is
delayed, the age of the stellar population will not coincide with the
formation age.  Fortunately, studies of globular clusters have revealed
that there do exist stars in our own Galaxy with ages that are nearly equal
to the age of the Universe (Marin-Franch \etal 2008; Salaris \& Weiss
2002).  Therefore, even if star formation is ongoing through a galaxy's
life, it may be possible to isolate the oldest stars and use their ages as
a lower limit to the age of the entire galaxy system and, by inference, the
epoch of galaxy formation.

In addition to a possible mismatch between the epoch of formation and the
epoch of initial star formation, there is also the possibility that the
stars in a galaxy do not form instantaneously as one unit.  For example, in
our own Galaxy there is a range of stellar ages (Twarog 1980), although the
halo population is uniformly old (De Angeli \etal 2005).  This problem can
be minimized by focusing our studies on galaxies where there is no evidence
of current or recent star formation.  Early-type galaxies satisfy this
condition with their lack of molecular gas and spectra profiles that are
dominated by evolved stars suggesting a majority are older than a few Gyr
(see Trager \etal 2005 for a dissenting view).

The age of a galaxy's stellar population has become an increasingly
important parameter in the last decade with the introduction of
hierarchical models of galaxy formation.  Under the original scenarios of
galaxy formation (referred to as the monolithic model, Larson 1974; Kodama
\& Arimoto 1997), a galaxy forms quickly, with an intense epoch of star
formation at high redshifts followed by a short phase of galactic winds.
The rapid initial star formation produces a stellar population that is
nearly singular in age and uniform in its chemical composition.  The epoch
of galactic winds is controlled by a galaxy's mass (i.e. depth of its
gravitational potential) and, thereby, its total metallicity.  Under
hierarchical models, the formation epochs are extended to lower redshifts,
assumingly with later epochs of star formation and thereby younger stellar
ages (White \& Frenk 1991).

The history of determining the age of a galaxy's stellar population is long
and rich in observational techniques (see Thomas \etal 2005 for a review).
Early studies focused on the direct comparison between integrated galaxy
colors and colors of galactic clusters (Sandage \& Vishvanathan 1977;
Burstein \etal 1987).  These early results supported a view where
early-type galaxy stellar populations were similar to globular clusters in
their evolutionary state, but with higher metallicities (Burstein \etal
1984).  Improved technology led to the matching of colors with various
spectral indices (Gonzalez 1993; Trager \etal 2000).  And improved
spectroenergy distribution (SED) models led to the use of spectral indices
alone to calculate the age and metallicity of underlying stellar
populations (Kuntschner 2000; Trager \etal 2000).

The method of determining age and metallicity of a galaxy matured with the
introduction of the Lick system that, primarily, depends on Fe (notably
Fe5270 and Fe5335), Mg $b$ and H$\beta$ lines to deduce mean age and
metallicity (Trager \etal 2000).  A surprising result from the
spectroscopic surveys (e.g., Gallazzi \etal 2006) is that a significant
fraction of early-type galaxies have mean ages younger than expected from
monolithic scenarios (see Schiavon 2007 for a review).  This result would
support hierarchical models of galaxy formation (Kauffmann, White \&
Guiderdoni 1993; Cowie \etal 1996).

A range of ages for early-type galaxies is not necessarily a problem for
their optical and near-IR colors as a stellar population's color evolution
is expected to proceed at a rapid pace for the first 1 to 2 Gyr after
initial star formation, but color changes over the next 10 Gyr are small
(Bower \etal 1992).  However, a significant number of early-type galaxies
with ages less than 8 Gyr would challenge our understanding of the thinness
of the Fundamental Plane (MacArthur \etal 2008; Cappellari \etal 2006,
where galaxies greater than $10^{11} M_{\sun}$ have formation redshifts of
greater than 2); the detection of evolved galaxies at high redshifts (Mei
\etal 2009, Andreon \etal 2008, where both studies find passive evolution
with redshifts of formation beyond 3); and the passive evolution of galaxies
at intermediate redshifts (Kelson \etal 2001, Rakos \& Schombert 1995,
where both studies find colors and spectral indices agree with passive
evolution models to redshifts of 0.8).

The goal of this paper is to examine the impact of young stellar population
age by comparison of expected colors for galaxies with spectroscopically
determined ages and metallicity against the color-magnitude relation (CMR).
Due to the known lack of uniqueness in age and metallicity for broadband
color systems (i.e. Johnson $UBV$ or SDSS $gri$, Worthey 1994), we compare
the SED model generated colors to a special narrowband color system that
focuses on spectral regions near the 4000\AA\ break.  To achieve this goal,
we have divided our analysis into three parts.  First, we examine the
behavior of the CMR in our color system and compare its slope and zeropoint
to other CMR's.  Second, we outline the selected SED models and their input
parameters.  Lastly, we construct CMR's from actual spectroscopic samples
and compare their colors with our observations.  Anticipating that the
derived colors, across all wavelengths, will be too blue to reproduce the
observed CMR, we will also examine the color-H$\beta$ phase space in order
to isolate the magnitude of the discrepancies.

\section{Color-Magnitude Relation}

Perhaps one of the oldest galaxy correlations is the one between galaxy
absolute luminosity and its global color (i.e., the color-magnitude
relation, Baum 1959; Faber 1973; Visvanathan \& Sandage 1977).  While first
discovered in near-UV and blue wavelengths (Caldwell 1983), subsequent
studies have demonstrated its existence from the far-UV (Kaviraj \etal
2007) to the near-IR (Chang \etal 2006).  All these studies can be
summarized such that 1) the correlation is one of redder galaxy colors with
increasing galaxy luminosity, 2) the correlation exists over a range of
galaxy environments, but strongest in the densest regions (Cooper \etal
2008), 3) the scatter in the correlation for early-type galaxies, while
small, is greater than observational errors (although most of the scatter
is associated with star-forming galaxies, Andreon 2003), 4) scatter for
early-type galaxies increases as the environment density decreases and 5)
there exist distinct red and blue components to the CMR, with the red
component being composed of early-type galaxies (Baldry \etal 2006).

\begin{figure}
\centering
\includegraphics[scale=0.95]{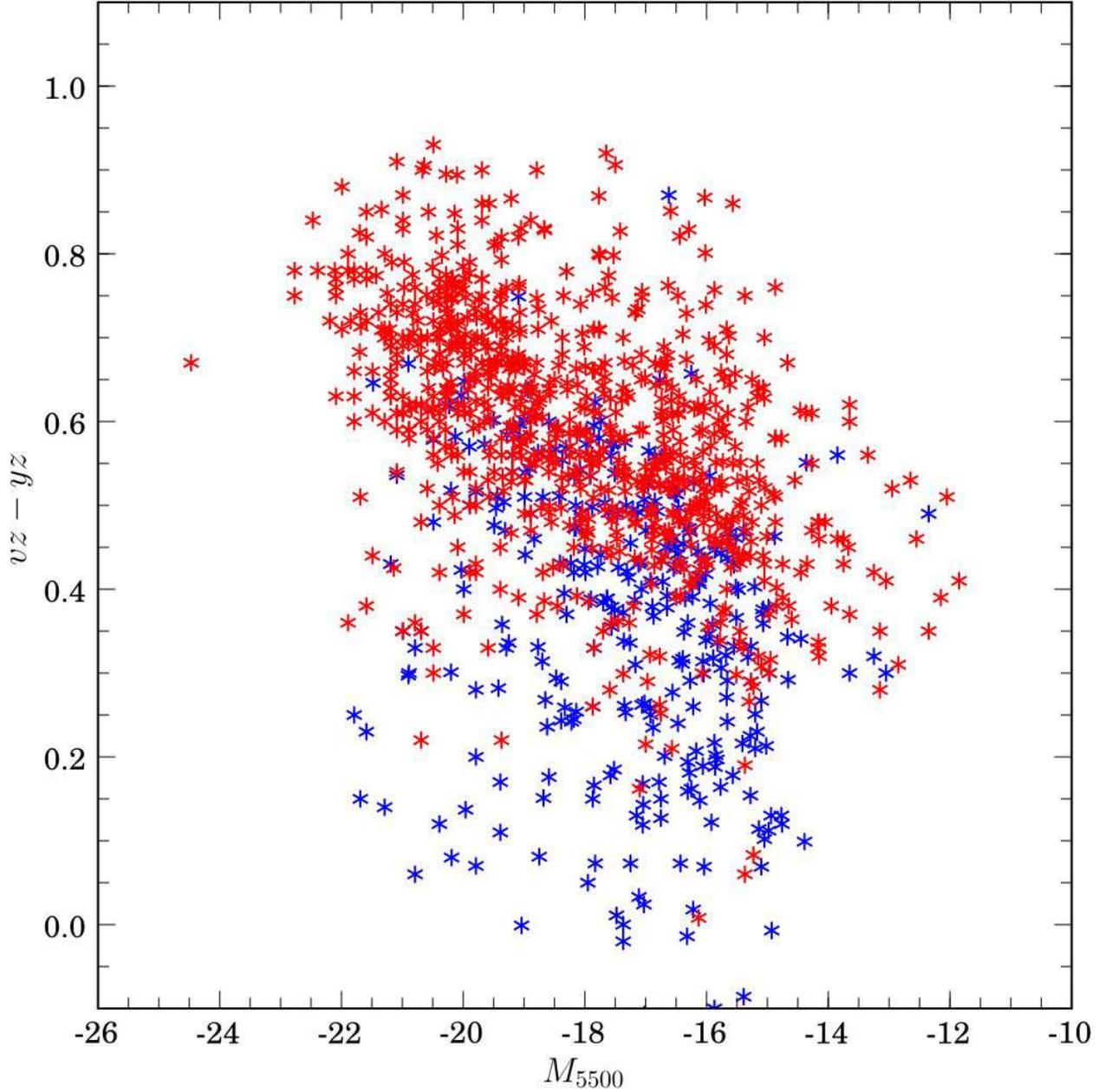}
\caption{The color-magnitude relation for 1,104 galaxies in our cluster
sample through our `metallicity color', $vz-yz$ (4100$\AA$ - 5500$\AA$).
The red and blue cluster populations are defined by division in our `continuum
color', $bz-yz$.  The red population ends up being 95\% early-type galaxies
by morphology and produces the strongest correlation between color and
luminosity of any filter combination (broadband or narrowband).
}
\end{figure}

Several direct corollaries are derived from the CMR with additional
knowledge concerning a galaxy's luminosity and color.  The first is that
galaxy luminosity is directly tied to the stellar mass of a galaxy
(Bernardi \etal 2003).  For early-type galaxies, this is the primary
component of their total baryonic mass.  The scaling factor from luminosity
to stellar mass is the $M/L$ ratio, that varies in a linear fashion from
2.5 to 5 for luminosities greater than $-$17 (Bernardi \etal 2003; Pahre,
Djorgovski \& de Carvalho 1998; Kelson \etal 2000) and is not a major
source of error in later derived relationships between mass and color.  The
conversion of luminosity to stellar mass is also confirmed by the behavior
of the color-$\sigma$ relation, the CMR's counterpart using a galaxy's
velocity dispersion instead of luminosity (Graves, Faber \& Schiavon 2008).

While a galaxy magnitude is directly related to stellar mass, the color of
a galaxy is a much more complicated observable to interpret.  For
early-type galaxies, a majority of a galaxy's spectrum is dominated by
light from the photospheres of the underlying stellar population; however,
different spectral regions are sensitive to different types of stars.  For
example, near-UV colors reflect the contribution of hot stars (Kaviraj
2007).  Typically, these are massive, young stars, but the contribution
from metal-poor HB stars or blue stragglers can not be ignored (Atlee,
Assef \& Kochanek 2008; Lisker \& Han 2008 ; Yi 2008).  The optical and
near-IR portions of the spectrum are strongly influenced by stars on the
RGB and, to a lesser extent, by turnoff stars (Pickles 1985; Rose 1985).
Where colors are dominated by the RGB, then changes in those colors reflect
changes in the mean effective temperature of the RGB stars.  Since changes
in age of RGB stars has very little effect on their temperatures (Charlot
\& Bruzual 1991), then changes in the color of the RGB is due, primarily,
to changes in mean metallicity (Matteucci 2006).

The dominance of the RGB to early-type galaxy colors was the primary reason
that the CMR, for this class of galaxies, was presumed to be a metallicity
effect (Faber 1973; Larson 1974).  Increases in the metallicity of a
stellar population lead to a decrease in RGB effective temperature (through
increased line blanketing) and, therefore, a redder integrated color.  In
later studies, Worthey, Trager \& Faber (1995) showed that an
age-luminosity correlation could also fit the CMR data due to an
age-metallicity degeneracy for broadband colors.  However, studies of high
redshift clusters (Kodama \& Arimoto 1997) discouraged an age
interpretation since the evolution of the CMR with redshift is less than
predicted by a passive evolution model (where it was assumed that a
majority of the stellar mass in a galaxy formed at high redshifts, i.e.
old, in a rapid burst, and later changed solely by stellar evolutionary
effects).

Regardless of whether the CMR is due to age and/or metallicity effects,
there is no doubt that the CMR (and its cousin, the color-$\sigma$
relation) relates the luminous and dynamical masses of a galaxy with the
properties of its underlying stellar populations.  And even evolution
models that attribute some of the CMR to the influence of age (younger
galaxies have lower luminosities), a majority of the color change must be
due to changes in metallicity (Thomas \etal 2005).  Higher metallicities
with higher luminosity (i.e., higher stellar or dynamic mass), as implied
by the CMR, can be produced by various galaxy formation scenarios.  Each of
these scenarios must result in the early cessation of star formation and/or
more inefficient chemical evolution for lower mass galaxies.  The typical
scenario (i.e. a classical wind model) requires rapid enrichment by initial
star formation with the chemical enrichment process halted by the removal
of a galaxy's ISM through SN driven galactic winds (Larson 1974, Matteucci
\& Tornambe 1987).

The difficulty in distinguishing age and metallicity effects on the CMR is
particularly salient for broadband filters, such as the Johnson $UBV$
system, as demonstrated by Worthey (1994).  In fact, under a rapid, single
initial burst scenario (the usually assumed scenario of galaxy formation
for early-type galaxies), broadband colors achieve stable values after only
3 to 4 Gyr (Bower \etal 1992).  This limits the ability of CMR to test age
versus metallicity effects and the traditional method of overcoming this
limitation is to extend the wavelength coverage, for example, using optical
to near-IR colors.  However, additional complications arise with near-IR
colors and the age-metallicity degeneracy concerning RGB and AGB ancestors of
high mass stars (Cordier \etal 2007).

An alternative approach, with respect to broadband colors, was pioneered by
Rakos \& Fialia (1983) with the use of optical narrowband filters
($\Delta\lambda = 200\AA$) based on the Str\"omgren system.  In this filter
system, two of the filters ($bz$ and $yz$) are located at 4675\AA\ and
5500\AA, regions of an early-type galaxy's spectrum that are relatively
free of metallicity features, and define a temperature or continuum color
index.  A third filter ($vz$) is located at 4100$\AA$.  This region is
shortward of the continuum at 4600\AA, but above the Balmer discontinuity,
and is strongly influenced by metal absorption lines (i.e. Fe, CN).  While
of limited value for young stellar populations, the $vz-yz$ color is very
sensitive to metallicity for spectral classes F to M, that dominate the
light in old stellar populations.  We parameterize the absolute magnitude
of a galaxy to the $M_{5500}$ luminosity where $m_{5500}$ is determined by
spectroscopic standards taken through our $yz$ filter (Rakos, Odell \&
Schombert 1997).  We have converted other studies $M_V$ or $M_B$ magnitudes
to $M_{5500}$ using their published $B-V$ colors (since the $V$ filter has
a slightly higher red side, the conversion from $M_V$ to $M_{5500}$ is such
that $M_{5500} = M_V + 0.005(B-V)$).

With respect to the CMR, a clear relationship has been demonstrated in all
the narrowband filters versus galaxy luminosity, and the resulting
correlations are stronger, and with less scatter, than the broadband versions.
For cluster early-type galaxies, Odell, Schombert \& Rakos (2002) outlined
the CMR over a broad range of galaxy absolute magnitudes ($M_{5500} = -$22
to $-$14) in Coma.  The strongest correlation (and steepest slope) is found
between $vz-yz$ color and luminosity.  This is in contrast to broadband
filters where the strongest correlations are found between near-UV colors
(i.e. $U-V$) and stronger than our own near-UV color index, $uz-vz$.  Our
early interpretation is that this confirms a strong metallicity influence
on the CMR, for if age was a significant contributor than the correlation
would be stronger in our continuum colors ($bz-yz$, Rakos \& Schombert
2005).

Since our work in Coma, we have sampled an additional eight rich clusters
out to redshifts of 0.17.  When combined with our previous work (over the
last 25 years), we have a total sample of 1,104 galaxies in the cores of
rich clusters (see Rakos \& Schombert 2008), and we present the CMR for all
1,104 galaxies in Figure 1.  In our previous studies, we have separated the
blue and red clusters populations based on a galaxy's $bz-yz$ color with
blue galaxies being defined as those with a $bz-yz$ index less than 0.22
(Rakos \etal 2000).  Division by color ties in a linear fashion with galaxy
morphology, where star-forming disk galaxies are easily excluded by this
selection criteria (see discussion in Rakos \& Schombert 2005b).  This
division by $bz-yz$ color is applied in Figure 1 and it is clear that the
red population forms a much stronger correlation with luminosity when the
blue population is excluded.

The fact that the CMR is a mixture of galaxies with different star
formation histories is well known (see Balogh \etal 2004).  This produces a
technical difficulty in using the CMR to interpret galaxy ages and
metallicities as any linear (or non-linear) fit to the data will be biased
towards the contaminating blue population.  Selection criteria using colors
introduces a reverse bias (Baldry \etal 2004) and selection by morphology
also introduces blue galaxies with early-type morphologies (Rakos \&
Schombert 2005b).  Recent studies using massive SDSS datasets can
deconvolve the red and blue components of the CMR (Balogh \etal 2002) and
indicates that the CMR of the red population is not linear with luminosity,
thus, making a linear fit to the data inappropriate.  This analysis is
consistent with the data for the red population in Figure 1.

Our method of choice for understanding the properties of the CMR in our
narrowband filters is to parameterize the luminosity/color space with a
ridgeline analysis.  This procedure divides the total sample in bins of
absolute luminosity.  Each bin's distribution of color is then subjected to
a gaussian fit to determine a peak color and normalized variance.  The peak
colors are used to define the ridgeline and this is a robust measure of a
maximum value since, if the number of data points is large, the bin size
can be varied as an internal measure of the accuracy of the mean color.

The calculated ridgeline for our $vz-yz$ color is shown in Figure 2.  Here
the red population (808 galaxies) are displayed greyscale contrast map.
This contrast map is produced by assuming that each data point is a 2D
gaussian of width given by its photometric error.  The individual gaussians
are summed and a greyscale intensity per pixel is assigned based on that
summation.  The pixel grid size is chosen to represent the mean error in
color.  The ridgeline follows the peak intensities of the greyscale plot,
as expected.  The CMR ridgeline is roughly linear to within the accuracies
of the method.  There is no strong indication of upward or downward trends
at the bright and faint ends, although there is the suggestion of a
flattening at the faint end (see also Baldry \etal 2004).

The linear shape of the ridgeline disagrees with the results of Baldry
\etal (2004) who find, even after separating the red and blue populations
in a 70,000 galaxy SDSS sample, that the $g-r$ CMR is not well fit by a
straight line, but rather a linear fit on the high luminosity end with a
tanh function at the fainter transition region ($M_{5500} = -19$).
Clearly, the difference in sample sizes makes the shape analysis of the
SDSS sample superior with our sample.  However, our sample is strictly a
cluster core sample, the highest galaxy density environment for any
evolutionary study.  There is every indication that environmental effects
vary across regions of differing density (Clemens \etal 2008) and across
galaxy mass (Hyeop Lee \etal 2008).  The more linear nature to our CMR may
simply reflect the homogeneous nature to evolutionary processes in cluster
cores.

\begin{figure}
\centering
\includegraphics[scale=0.95]{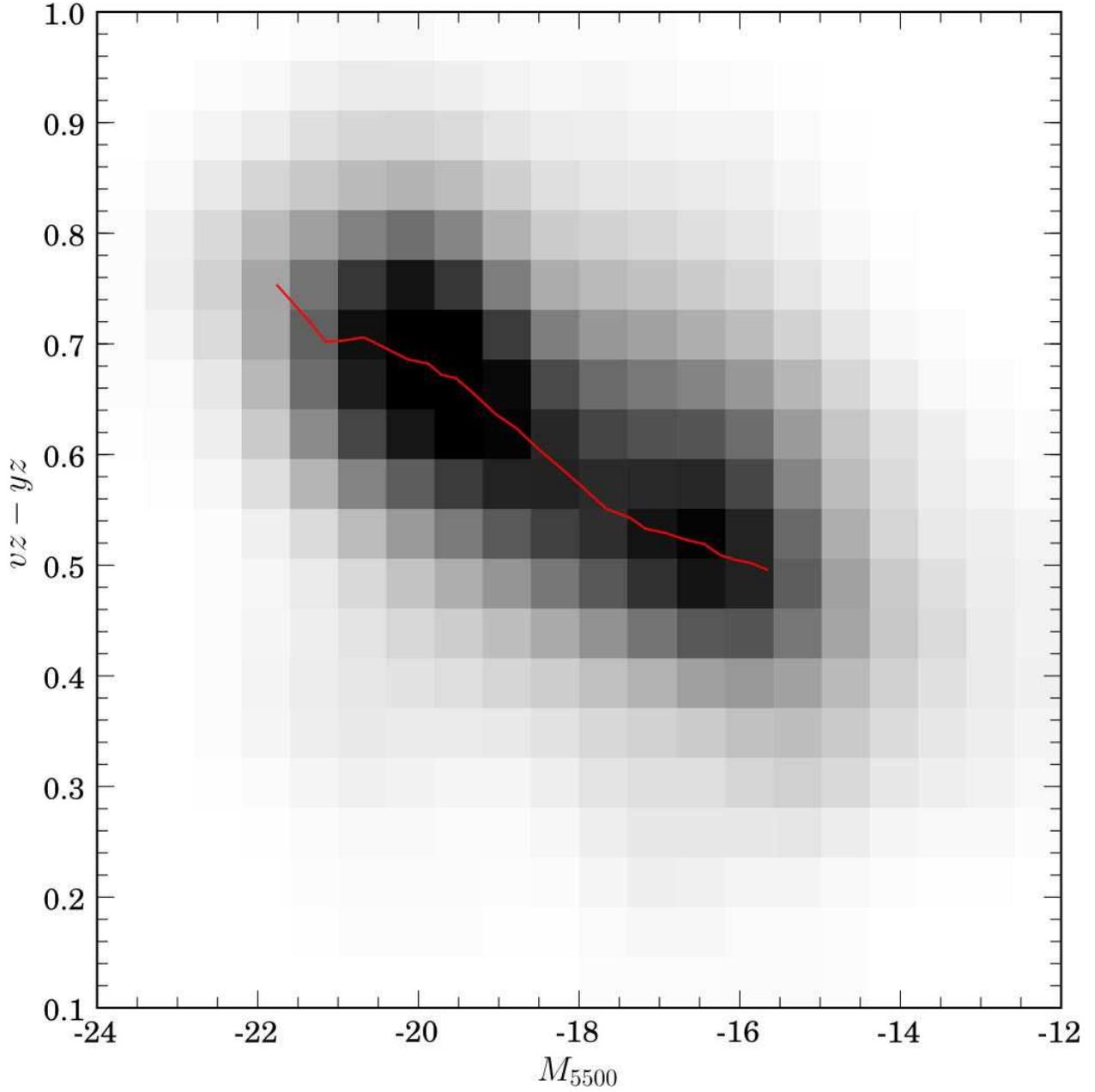}
\caption{A contrast map of the $vz-yz$ CMR.  Each individual data point
is treated as a 2D gaussian with its width determined by the photometric
errors.  All the individual gaussians are summed and binned to produce a
greyscale map where intensity represents the data density at that pixel
value.  The red line represents the ridgeline defined by the gaussian data.
The ridgeline is linear to within the errors.
}
\end{figure}

To summarize our results on the $vz-yz$ CMR, we found the following
discernible correlations that are relevant to our comparison of a galaxy's
age and metallicity.  First, there are clearly two populations involved
when a galaxy's absolute luminosity is compared with its color index.  A red
population, whose CMR is increasing in color with luminosity and whose
colors are indicative of a non-star-forming system, plus a blue population,
whose colors indicate recent (last few Gyr) or ongoing star formation.  A
ridgeline analysis of the red population displays a well correlated
relationship between luminosity and color that is, approximately, linear to
within the internal errors.  The CMR is steepest in our $vz-yz$ colors,
compared with $uz-vz$ or $bz-yz$, which is slightly surprising as the CMR is
strongest in the bluest colors for the Johnson $UBV$ system (Wyder \etal
2007).  Since $vz-yz$ is a color sensitive to metallicity effects, this
will support a metallicity interpretation to the CMR.

\section{Interpretation of the CMR}

In order to interpret the colors of a galaxy, as they reflect its star
formation history, one must consult spectroevolutionary (SED) models (see
Schiavon 2007 for a review).  These models gather all the relevant star
formation information (the assumed IMF, age, metallicity, star formation
rate) and output a summed spectrum of the entire underlying stellar
population.  These spectra can then be convolved through our various
filters to produce the narrowband colors that we observe.

In recent years there have been a number of SED models in the literature
(Cid Fernandes \etal 2008; Franzetti \etal 2008; Li \& Han 2007; Schulz
\etal 2002).  With respect to our narrowband indices, we have found that
most have converged to identical results for stellar populations older than
1 Gyr (see Schombert \& Rakos 2009 for a full discussion of the various
models and the predicted colors).  For this study, we have selected the
Bruzual \& Charlot (2003, hereafter, BC03) models, although the choice of a
particular flavor of models does not change our results.

The BC03 models provide the user with a range of ages (from 0.1 to 20 Gyr)
and metallicities (from $-$2.3 to $+$0.4 [Fe/H]) for a simple stellar
population (SSP), i.e. a stellar population that is singular in age and
metallicity.  For our purposes, we have selected a fairly standard range of
models from 1 to 14 Gyr and $-$2.3 to $+$0.4 in [Fe/H], all using the
Chabrier (2003) IMF (mass cutoff at 0.1 and 100 $M_{\sun}$) and Padova 2000
isochrones.  Each SSP is interpolated at the 0.1 dex level in metallicity
and convolved to our narrowband filters to produce a full grid of colors.
Table 1 lists the SSP colors (our narrowband system, Johnson $U-B$, $B-V$,
$V-K$ and the Lick indices $<$Fe$>$, Mg$b$ and H$\beta$) for two ages, 5 and
12 Gyr.  The reddest (i.e. most massive) ellipticals ($vz-yz=0.75$ and
$B-V=0.95$) would have higher than solar metallicities for 5 Gyr models and
slightly less than solar metallicities for 12 Gyr models.

\begin{figure}
\centering
\includegraphics[scale=0.95]{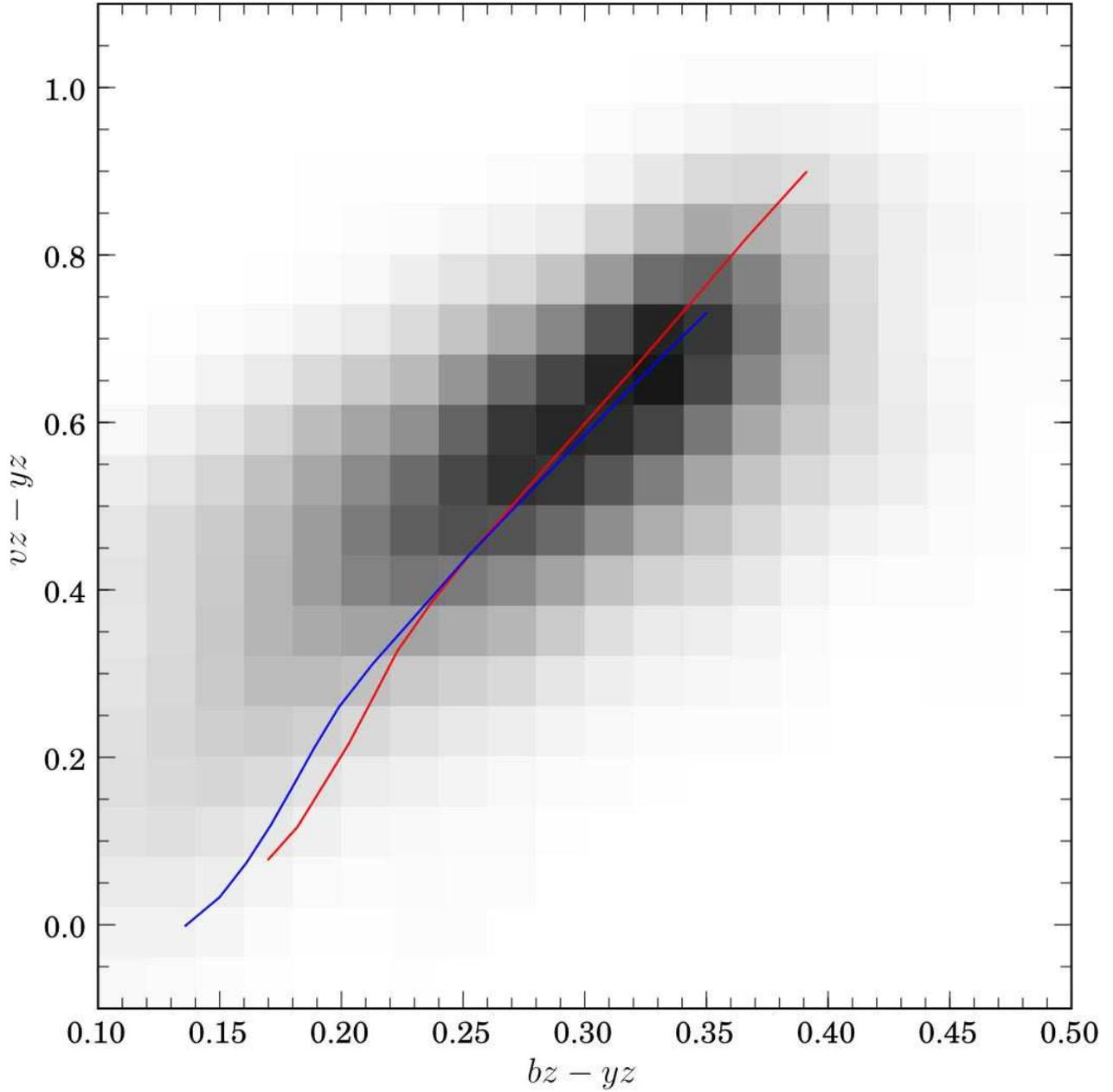}
\caption{The two color diagram, our metallicity
color, $vz-yz$, versus our continuum color, $bz-yz$.  The correlation
between colors is well known (Rakos \& Schombert 2007).  The red line is a 12 Gyr
multi-metallicity model (see discussion), the blue line is a 5 Gyr model.
Either is an adequate description of galaxy colors, although the younger
models require super-solar metallicities for the reddest ellipticals.
}
\end{figure}

A more advanced use of SED models is assume that galaxies are not composed
solely of SSPs, but rather a sum of SSP's of varying metallicities
following a simple chemical evolution model (Schombert \& Rakos 2009).
These chemical evolution models can take on a range of shapes, as a
function of [Fe/H], dependent on initial conditions and physics input to
the models.  We have used a simple infall scenario (Kodama \& Arimoto 1997)
that matches the shape of the internal metallicity distributions for the
Milky Way (Wyse \& Gilmore 1995), M31 (Worthey \etal 2005) and NGC 5128
(Harris \& Harris 2000).  We have then allowed this metallicity
distribution to slide in the peak [Fe/H] value to produce a range total
metallicities per galaxy age (see our `push' model, Schombert \& Rakos
2009).  We argue in that study that this model matches the $<$Fe$>$ versus
color plane but, again, the exact shape of the adopted metallicity
distribution is not a key parameter to our following conclusions.

Each model produces a metallicity value based on a sum of the model
metallicity bins (a numerical average) and a luminosity weighted value
(metal-poor stars are hotter, and therefore brighter, than metal-rich
stars).  The difference between these values has a minor correction as
outlined in Schombert \& Rakos (2009), in other words, it is easy to
convert between the two values.  In our following analysis, we always use
the luminosity-weighted [Fe/H] values.  Examples of our multi-metallicity
model results are listed in Table 1.  Our 5 and 12 Gyr model (for [Fe/H]
values from $-$2 to $+$0.8) is shown in Figure 3, a two color diagram
($vz-yz$ versus $bz-yz$) for our cluster sample.  The 12 Gyr model agrees
well with the ridgeline of the data; however, a large range of ages would
equally satisfy the data and this diagram is a poor indicator of mean age
and metallicity (Rakos \& Schombert 2004).  We present this plot to
demonstrate the models span the same range of color space as the
observations.  We also note that the accuracy of the models is limited by
the accuracy of the input age and metallicity values.  Typical errors
quoted for age and [Fe/H] in the literature are 0.5 Gyr and 0.2 dex (Rakos
\& Schombert 2005).  These variations map into a model color error for
$vz-yz$ of 0.01 and 0.06 respectfully.  These are similar to the
photometric errors for the faint end of our cluster samples and, thus, we
do not expect to be model limited in our interpretations.

There are several features to note in Table 1 with respect to comparing our
multi-metallicity models with SSP's.  First, for the same mean [Fe/H], the
multi-metallicity models produce bluer colors ($-$0.13 in $\Delta(vz-yz)$,
$-$0.04 in $\Delta(B-V)$, $-$0.05 in $\Delta(V-K)$) than their equivalent
SSP models.  This is due, of course, to the inclusion of a metal-poor tail
that contributes more strongly to color than mean [Fe/H].  For our
multi-metallicity models, red ellipticals would only have solar and higher
metallicity for ages greater than 10 Gyr.  Younger models would require
super-solar metallicities to explain the colors of the brightest
ellipticals, values that are inconsistent with $<$Fe$>$ values (between 2.3
and 3.0).  Whereas, the older multi-metallicity values have metallicity
indices, $<$Fe$>$ and Mg$b$, in agreement with the colors of bright
ellipticals.  High H$\beta$ values, i.e. greater than 2, are a challenge
for older multi-metallicity models as they predict extremely blue colors.
This will be discussed further in \S5.

Lastly, we need to consider the effects of $\alpha$-enhancement to the SED
models.  It has been known for some time that massive early-type galaxies
are enhanced in light elements (so-called $\alpha$ elements, see Worthey
\etal 1992) compared with Fe.  SED models that use varying ratios of
$\alpha$/Fe will predict differing colors, mostly due to variations in the
effective temperature of the RGB.  Thomas \etal (2005) find [$\alpha$/Fe]
varies from 0 to 0.3 in cluster ellipticals.  This agrees well with Trager
\etal (2008) Mg/Fe ratios in Coma.  For our models, we consider the changes
in color ($B-V$) presented for the [$\alpha$/Fe]=+0.3 from Schiavon (2007)
and Percival \etal (2009).  For those models, with ages greater than 5
Gyrs, it is found that $B-V$ reddens by 0.02 between the solar and
$\alpha$-enhanced models.  This translates into $\Delta(vz-yz) = 0.03$ for
our color system, and will be our baseline for comparison in \S4.  As the
$vz$ filter is centered on a region of the spectrum (4600\AA) rich in
metallicity lines, it is possible that $\alpha$ enhanced models will
deviate $vz-yz$ redward to a greater degree than $B-V$.  However, as most
of the differences are due to RGB temperatures, not line blanketing
effects, we assume that a change of 0.03 is a reasonable approximation for
$\alpha$/Fe variations.

While our narrowband color system is an improvement over broadband colors
in discriminating age and metallicity (see Rakos \& Schombert 2005 and
Rakos \& Schombert 2007), there is still a degeneracy in going from color
to age and metallicity.  However, it is the nature of the underlying
stellar population that the two parameters of age and metallicity define a
unique color.  Thus, we can use the SED models to deduce the expected
colors of galaxy's with age and metallicity values determined in another
fashion, in this case, spectroscopically.

\section{Spectroscopic Ages}

Attempting to extract age and metallicity from galaxy colors is a difficult
procedure with a great deal of internal error associated with the well
known age-metallicity degeneracy problems (see Rakos \& Schombert 2005).
However, in order to test the validity of spectroscopic age and metallicity
determinations, we need only use the observed spectral index age and
metallicity values and calculate their expected narrowband colors.  Where a
set of galaxy colors may be due to a range of ages and metallicities, a
couplet of age and metallicity values has only a singular set of
corresponding colors.  Thus, this procedure depends only on the accuracy of
the SED models and produces a unique set of colors without degeneracy.

Using the multi-metallicity SED models from \S3, and the luminosities, ages
and metallicities listed in Trager \etal (2008) and Thomas \etal (2005), we
can calculate the expected colors and plot these values on the $vz-yz$ CMR
(a galaxy's total luminosity is independent of the models).  This
experiment is shown in Figure 4, where the blue symbols are from the Trager
\etal sample and the green symbols are from the Thomas \etal sample.  The
greyscale image is our cluster data and the red ridgeline taken from Figure
2.  It is immediately obvious that neither dataset matches the CMR for
cluster galaxies.  For the combined Trager \etal and Thomas \etal datasets,
97\% of their galaxies lie below (i.e., having bluer colors) the ridgeline.
While the predicted colors are not outside the range of galaxy colors for
the entire sample (e.g., the lower luminosity galaxies), they are
significantly bluer than galaxies of similar absolute luminosity.  Both
Trager \etal and Thomas \etal samples are dominated by high mass galaxies
($M_{5500} < -19$), but there is no evidence that colors for the low mass
end of the sample are closer to the CMR then those on the high mass end.

\begin{figure}
\centering
\includegraphics[scale=0.95]{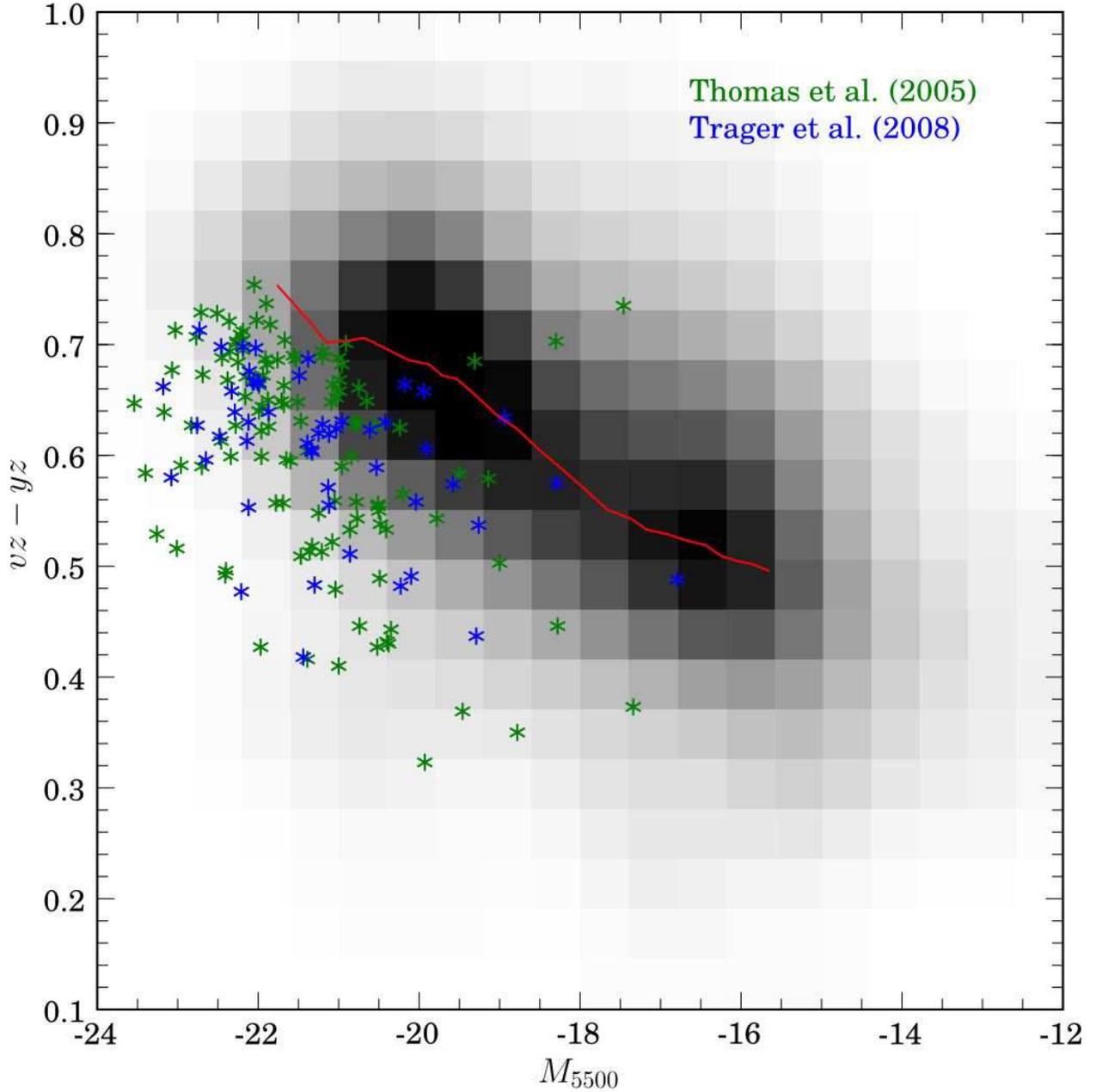}
\caption{The calculated $vz-yz$ colors for the samples of Thomas \etal
(2005) (green symbols) and Trager \etal (2008) (blue symbols).  The
spectroscopic ages and metallicities from those studies are used to
calculate their $vz-yz$, and then plotted against $M_{5500}$ absolute
magnitude.  Our $vz-yz$ cluster data and ridgeline (red line) from Figure 2
are also shown.  A majority of the spectroscopic data lie blueward of the
$vz-yz$ CMR indicating that the spectroscopically measured age and/or
metallicity values are in error.
}
\end{figure}

The magnitude of the difference between the spectroscopic colors and the
CMR, as a function of absolute luminosity, is shown in Figure 5.  Shown in
this Figure is the difference between the model color (from the
spectroscopically determined age and metallicity) and the CMR ridgeline
(the scatter around the ridgeline is 0.08).  The dotted red line is the
mean color difference for the sum of all three samples, which is 0.12 mags
blueward of the CMR.  There is no correlation between color separation from
the CMR and luminosity (stellar mass).  Also shown are the errors in model
color due to errors in the age and metallicity estimates.  Neither error is
sufficient to explain the discrepancy in color.  Also shown is the color
effect of assuming a different $\alpha$/Fe ratio (which reddens color for a
given age and metallicity, see discussion in the previous section).

As a secondary check to our procedures, over a larger range in luminosity,
we have taken the linear relationships between galaxy mass, age and
metallicity from the early-type Gallazzi \etal (2006) SDSS sample (see
their Table 4) and calculated $vz-yz$ colors for that set of ages and
metallicities.  Although these relations, quoted in Gallazzi \etal Table 4
are linear, they, in fact, predict increasing [Fe/H] with increasing
age, in contradiction with their Figure 18.  This simply reflects the spread
in [Fe/H] per stellar mass bin (see their Figure 17).  To better represent
the SDSS dataset, we have outlined a region in color space that marks the
boundaries for [Fe/H] (from their Figure 17, an [Fe/H] range of $-$0.2 to
+0.1 at a mass of 10$^{9.75}$ and a range of 0.0 to +0.3 at a mass of
10$^{11.5}$) and adopts the linear relation of age and stellar mass.  To
convert stellar mass into $M_{5500}$ we have adopted the relationship
between $M_r$ and $M_*$ from Table 4 of Gallazzi \etal (2006) ($M_r = -2.29
{\rm log} M_* + 3.45$) and assuming $V-r=0.30$ from SDSS transforms for a
mean color of $g-r=0.75$.

\begin{figure}
\centering
\includegraphics[scale=0.95]{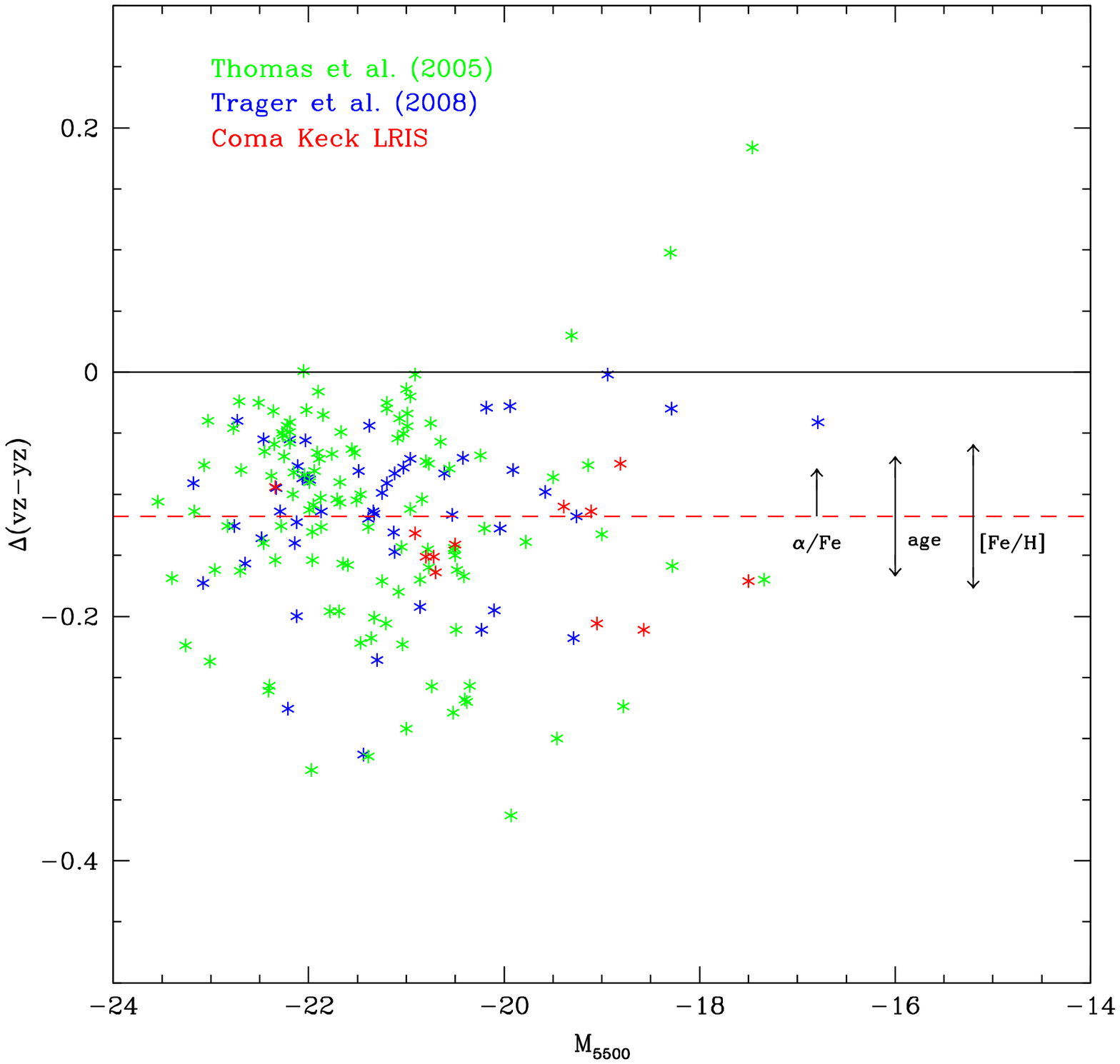}
\caption{The difference between the model colors (as given by spectroscopic
age and metallicity) and the CMR ridgeline from Figure 4.  Both Trager
\etal and Thomas \etal samples are shown, as well as the high S/N Keck
sample (Trager \etal 2008) of bright galaxies in Coma.  The red dotted line
is the mean color difference for the samples.  The age and [Fe/H] arrows
display the spectroscopic errors for age and metallicity as they reflect
into model color.  Also shown is the shift in color needed to correct for
enhanced $\alpha$/Fe models.  None of the errors or corrections approach
values needed to explain away the discrepancy with the CMR.
}
\end{figure}

That experiment is shown in Figure 6 where the relationships from Gallazzi
\etal age and metallicities are converted into $vz-yz$ colors.  As with the
Trager \etal and Thomas \etal samples, the Gallazzi SDSS sample also lies
below the CMR ridgeline, a closer match at the low mass end where galaxies
have spectroscopic ages of 7 Gyrs and mean [Fe/H] of near solar.  The
colors of the high mass end are bluer than the observed CMR, although not
as dramatic as the Thomas \etal and Trager \etal samples.  This problem is
also noted if we use their age and metallicity to predict SDSS colors
($g-r$) (see Gallazzi \etal (2006), Figure 4) or Johnson $UBV$ colors.  We
also note that applying any correction for $\alpha$-enhancement would only
redden the predicted colors by 0.03 (see discussion in previous section),
which is insufficient to explain the discrepancies in Figures 5 or 6.

It only remains to determine which parameter, age or metallicity, is
responsible for the deviations from the CMR.  The obvious candidate is
stellar population age since the difference is in the direction of bluer
colors, and ages less than 8 Gyr have a larger impact on optical color than
metallicity.  In addition, the metallicities quoted by all three
spectroscopic studies are in fair agreement with direct metallicity
measurements based on high resolution imaging of nearby galaxy's CMD's
(Worthey \etal 2005).  In order to redden the colors of either the Trager
\etal, Thomas \etal or Gallazzi \etal samples using solely changes in
metallicity (in order to maintain their young ages) would require
increasing their mean [Fe/H] to super-solar values, in conflict with the
metallicity of the Milky Way and other Local Group galaxies where
metallicity is determined directly through CMD diagrams.  Ages of 5 to 7 Gyr
for low mass ($10^9 M_{\sun}$) galaxies can be preserved, but require solar
metallicity values, in contradiction with the spectroscopic studies own
[Fe/H] estimates (although the SDSS samples display a large range in
metallicity at the low mass end).

\begin{figure}
\centering
\includegraphics[scale=0.95]{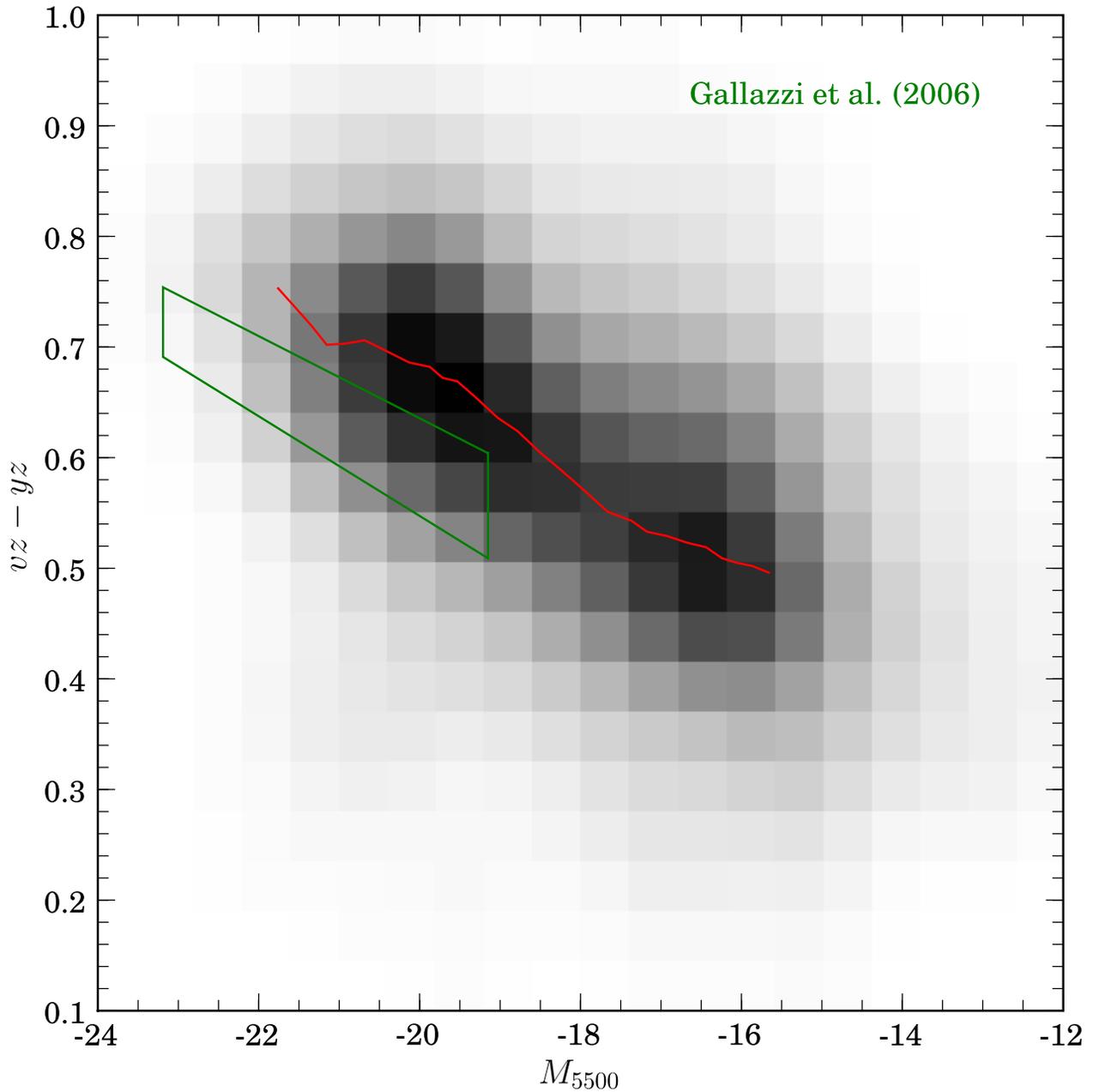}
\caption{The calculated $vz-yz$ colors for the SDSS sample of Gallazzi
\etal (2006).  We have used their derived 
bounds for age and metallicity (their Table 4, green polygon) to calculate
$vz-yz$ color.  As with Thomas \etal and Trager \etal samples, the SDSS
mean values all lie blueward of the mean CMR (red line), although
achieve a better match as low masses.
}
\end{figure}

In order to test the hypothesis that age is primarily responsible for the
deviations in Figures 5 and 6, we have re-calculated all the colors for the
three samples using the spectroscopic metallicity values, but assuming a
mean age of 12 Gyr rather than the spectroscopically determined age.  We
note that since age and metallicity are determined from the H$\beta$/[MgFe]
diagram, and lines of constant metallicity are nearly vertical in this
diagram, there is only a small correction to the calculated metallicity
from the MgFe index if a galaxy's age is increased.  This correction is in
the direction of decreasing the total metallicity (bluer colors) and has
the magnitude of approximately 0.2 dex from 12 to 7 Gyr.  Interestingly,
this correction is balanced by difference between a metallicity determined
by SSP's versus a luminosity weighted value for [Fe/H] from a stellar
population that has a range of metallicities (this correction increases the
mean [Fe/H] by 0.25 dex, see Schombert \& Rakos 2009).  This balance may
explain why the spectroscopic [Fe/H] values for galaxies agrees with other
techniques, while underestimating a galaxy's mean age.

The age corrected data points are shown in Figure 7.  Using an old stellar
age, we find that both the Trager \etal and Thomas \etal data are well
matched to the CMR, and the slope of the CMR is completely determined by
changes solely in mean metallicity with galaxy mass.  The excellent match
for the corrected data is slightly deceiving as it is assisted by the fact
that there is a correlation between age and metallicity in the Thomas \etal
and Trager \etal samples in the sense that the oldest galaxies have the
lowest metallicities.  Their reddest galaxies (with ages greater than 12
Gyrs) have the lowest metallicity and lie only slightly below the ridgeline.
Correcting these colors to 12 Gyrs moves them only slightly bluer.
Galaxies with the youngest ages have the highest metallicity, and
correcting their colors to a 12 Gyr age places them above the ridgeline (on
average) due to their higher [Fe/H] values.  However, the mean effect for
the corrected sample is valid, correcting young ages to 12 Gyrs reddens
their colors by an amount that places them in agreement with the CMR.

In addition, we have corrected the Gallazzi \etal relationship to 12 Gyr,
also shown in Figure 7.  The resulting region is an excellent match to the
CMR ridgeline in both $vz-yz$ and $g-r$ (not shown), at the low mass end,
but slightly flatter than the observed CMR.  The Gallazzi \etal region also
nicely brackets the corrected Trager \etal and Thomas \etal samples.  The
corrected region is slightly lower for the high mass end ($M_{5500} <
-22$).  To within the accuracy of the measurements, and our technique, we
find that high luminosity early-type galaxy's color can be reproduced by
assuming old ages ($\tau > 12$ Gyr).  In addition, the corrected colors
match the CMR such that there is no evidence of a significant
population of galaxies younger than 10 Gyr, at least for luminosities
greater than $-$19.

\begin{figure}
\centering
\includegraphics[scale=0.95]{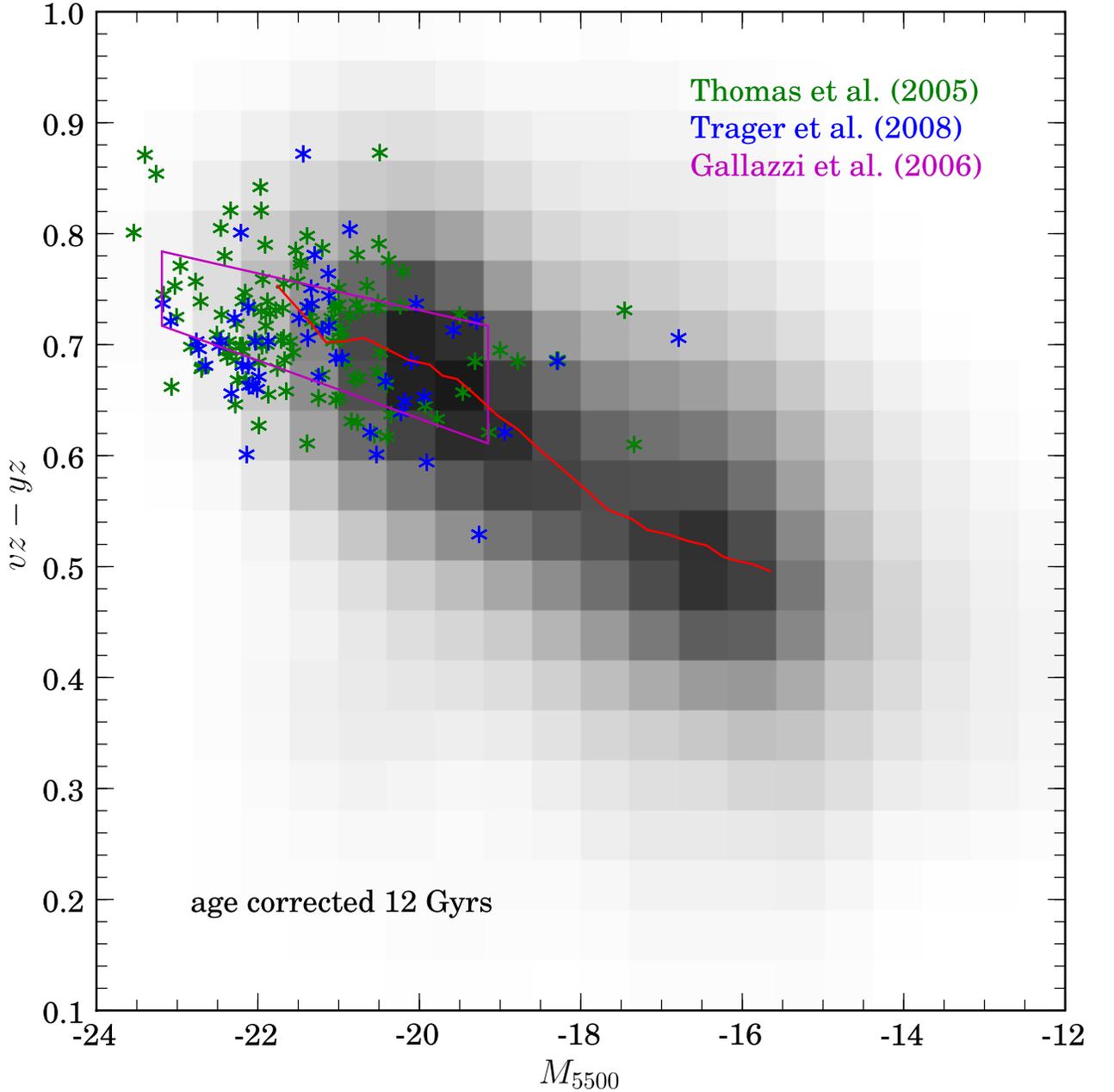}
\caption{The age corrected CMR for all three spectroscopic samples.  These
are the same samples as Figures 5 and 6; however, the galaxy colors are
calculated using an assumed age of 12 Gyr rather than the
spectroscopically determined ages.  The spectroscopically determined
metallicity values are used, although an older age would imply slightly
lower [Fe/H] values as calculated from metallicity indices such as MgFe.
All three spectroscopic samples are now in agreement with the CMR.
}
\end{figure}

As a last check to our technique we have examined the high resolution data
from the Coma cluster core galaxy by Trager \etal (2008).  This dataset
represents the highest quality spectroscopic indices to date and is
analyzed to dismiss any debate that spectroscopic ages and metallicities
are inherently inaccurate.   The resulting age and metallicities for the
sample are listed in Table 5 of Trager \etal (2008).  They range from 3.0
to 9.2 Gyr and $-$0.25 to $+$0.54 in [Fe/H].  Narrowband colors were
obtained for these galaxies in Odell, Schombert \& Rakos (2002).  Model
colors ($vz-yz_{model}$) versus observed colors are shown in Figure 8,
where we generated model colors using the Trager \etal (2008) ages and
metallicities (black labels).  Again, we see that the predicted model
colors are bluer than the observed colors.  And, again, we calculate new
model colors using an age of 12 Gyr and [Fe/H] values determined strictly
from the $<$Fe$>$ index (Schombert \& Rakos 2009).  These values are shown
as red symbols in Figure 8, and as our previous results we find the new
colors in excellent agreement with the observed colors (the one-to-one line
is shown in blue for Figure 8).

We can reverse our previous arguments with the Coma data by holding the
metallicity fixed at some particular value (e.g. solar) and use the
spectroscopic ages to determine the colors.  This experiment is shown in
Figure 8 as the green symbols where all the metallicities were assumed to
be solar.  The metallicity data is a poor match to the observations, and a
fixed metallicity is clearly wrong given the variations in $<$Fe$>$.
However, if one uses the spectroscopic ages, and increases the [Fe/H] value
for each galaxy to match the observed colors, then all the Coma galaxies
require a mean increase in [Fe/H] of $+$0.5 dex, ranging from solar for the
low mass galaxies to $+$0.8.  However, these values disagree with other
metallicity indicators in cluster galaxies (Graves \& Schiavon 2008).

\begin{figure}
\centering
\includegraphics[scale=0.95]{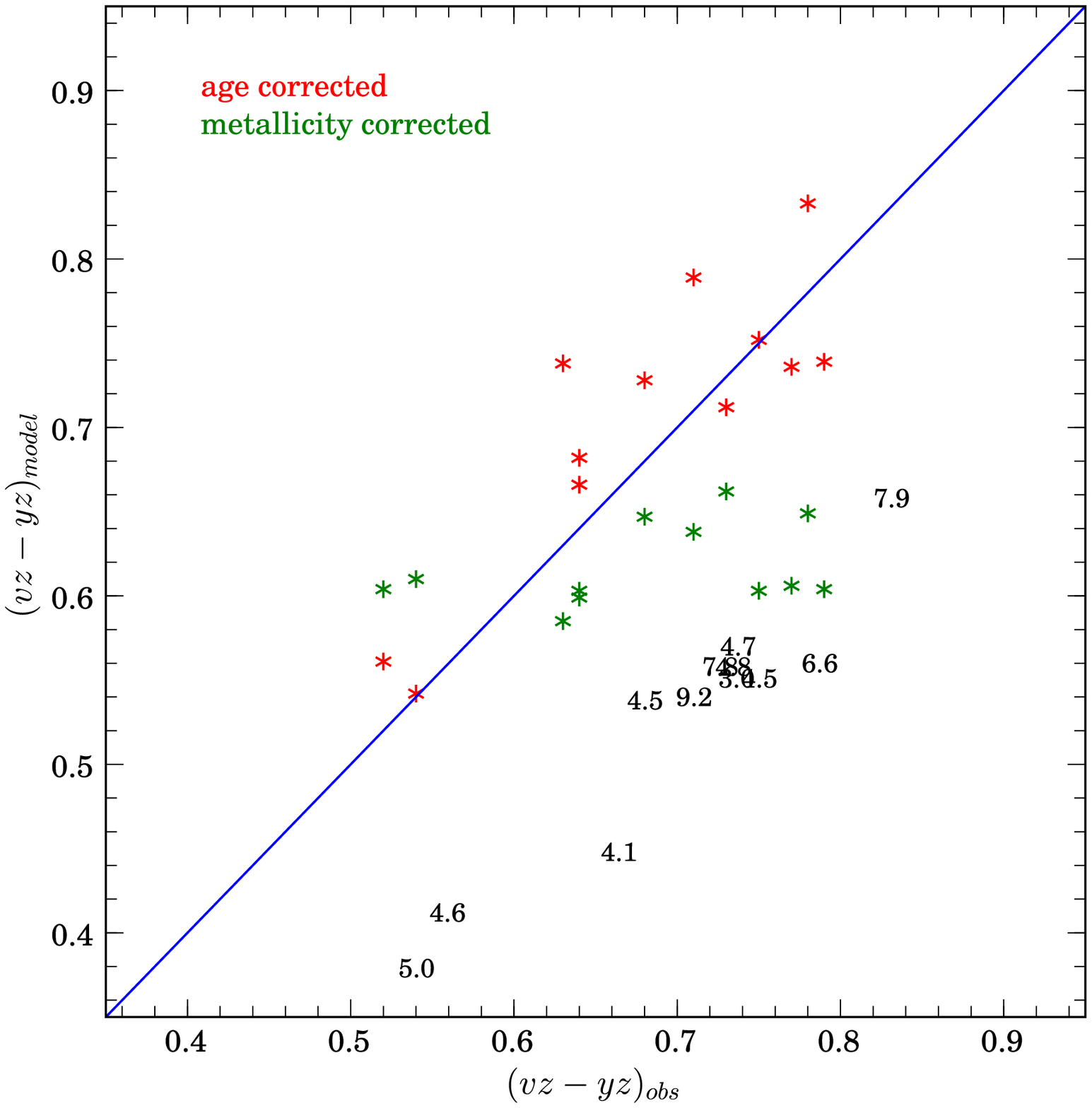}
\caption{The Coma Keck spectroscopic sample from Trager \etal (2008).  This
sample represents the highest quality spectral indices for age and
metallicity determination to date.  The black symbols are the observed narrowband
$vz-yz$ colors (Odell, Schombert \& Rakos 2002) versus their model colors
using spectroscopic age and metallicity values.  The galaxies spectroscopic
ages are shown.  The red symbols display the same galaxies with model
colors assuming an age of 12 Gyr and metallicity determined solely from the
$<$Fe$>$ index.  The green symbols represent the same analysis, only using
the spectroscopic ages, but fixing the metallicity to solar values.
}
\end{figure}

Our analysis in Figure 8 differs from our previous technique by using the
$<$Fe$>$ to determine [Fe/H] in a model independent fashion.  This leaves
age as the sole independent parameter and, again, an old stellar age is
required to reproduce the observed colors.  The importance of the Trager
\etal Keck data is that it is clear that the younger age estimates from the
spectroscopic data are driven by high H$\beta$ values for the data is of
such a high quality there is little uncertainty to their measured values.
And there can be no doubt from the Keck data that the H$\beta$ values for
all the Coma galaxies are greater than expected for any standard model of a
12 Gyr population (see Table 1).  Therefore, the dilemma for any SED model
is to reproduce the H$\beta$ values measured by numerous spectroscopic
surveys and SDSS samples, yet maintain red continuum colors (from the
near-UV to the near-IR) for the galaxy as a whole.

\section{Narrowband Colors and H$\beta$}

While all indications are that age is primarily responsible for problems
with the spectroscopic samples, there is no clear understanding for why
this is the case.  A stellar population's age is determined, primarily,
from its H$\beta$ index.  The spectroscopic studies considered herein
deduce age from the MgFe versus H$\beta$ diagram using similar models from
which we extract our narrowband colors.  In fact, for the Coma sample, one
can use the MgFe and H$\beta$ values to reproduce the metallicity and ages
in Trager \etal (2008), so model interpretation is not the central problem.
In particular, even for multi-metallicity models (Schombert \& Rakos 2009),
old galaxies have H$\beta$ values around 1.4, while a majority of normal,
early-type galaxies have H$\beta$ greater than 1.4 (Cervantes \& Vazdekis
2008), indicative of ages less than 5 Gyr from the same SED models.

\begin{figure}
\centering
\includegraphics[scale=0.95]{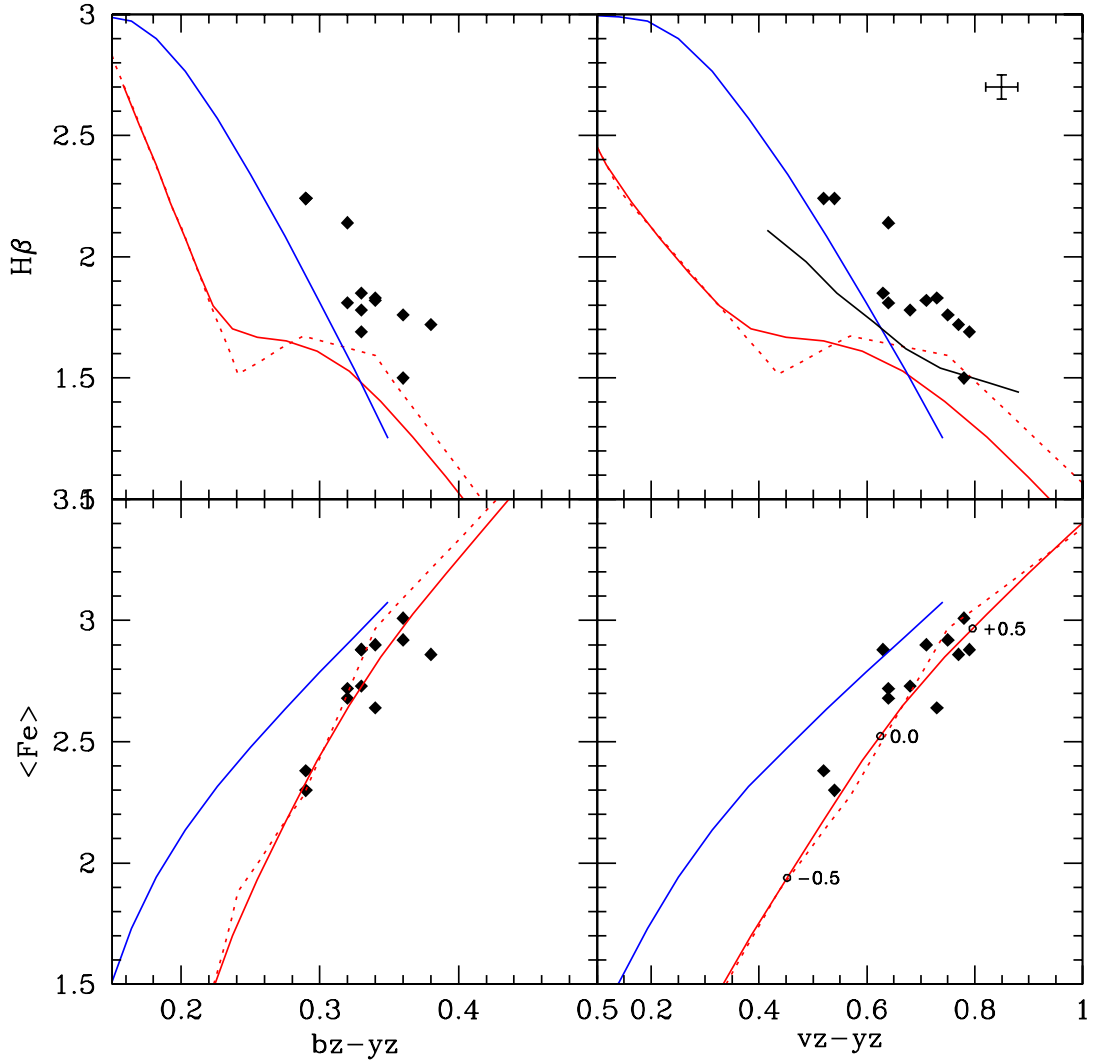}
\caption{The Trager \etal (2008) Coma data combined with our metallicity
($vz-yz$) and continuum ($bz-yz$) colors.  The solid red line is our
multi-metallicity 12 Gyr model ([Fe/H] values are indicated), the dashed
red line is a Bruzual \& Charlot 12 Gyr SSP.  The blue line is a 'frosting'
model, a 10\% 1 Gyr population added to the 12 Gyr model.  None of the
models adequately match the H$\beta$ values; however, the multi-metallicity
12 Gyr model is an excellent fit to the $<$Fe$>$ versus color diagram,
superior to the SSP models.  The black solid line represents the
$\alpha$-enhanced models of Idiart \etal (2007).
}
\end{figure}

There are several scenarios to increase the H$\beta$ signature in an old
population.  However, all of them require the inclusion of a hot stellar
component that results in bluer colors, both in broadband and narrowband
systems.  For example, increasing the number of metal-poor stars in a
galaxy will increase the H$\beta$ index through increases in temperature of
the RGB.  A first order estimate, based on numerical experiments with SED
models, finds a change in H$\beta$ from 1.4 to 1.6 produces a decrease
(bluer) $vz-yz$ color of 0.16 based on a 10\% increase in the number of
metal-poor stars ([Fe/H] $< -1.5$).  This shift would be in disagreement
with the mean colors of the CMR.

Other exotic populations with strong H$\beta$ signatures (blue stragglers,
blue horizontal branch stars, hot white dwarfs) all suffer from the
same effect of bluer continuum colors with shifts between $-$0.25 and
$-$0.40 to the blue for changes of only 0.2 in H$\beta$.  None of these
changes in color are consistent with the CMR.

Introducing young, massive stars is slightly more successful than the
previous two experiments.  This technique involves mixing a younger stellar
population with an older (12 Gyr) population, referred to as
`frosting' models (Trager \etal 2000).  For our analysis we have mixed a 1
Gyr population with our multi-metallicity models of 12 Gyrs.  Figure 9
displays the Trager \etal (2008) Coma data with our 12 Gyr model and a 10\%
frosting model (note: this mixture results in a spectroscopic age of
approximately 2 Gyrs), plotting our metallicity ($vz-yz$) and continuum color
($bz-yz$) versus the H$\beta$ and $<$Fe$>$ indices.  The 12 Gyr models are
an excellent fit to our colors versus $<$Fe$>$; however, fail to match the
H$\beta$ values.  The frosting models increases the H$\beta$ values, but,
as expected, also decrease the colors and fail to match the data.  It is
instructive to note that the multi-metallicity 12 Gyr model matches the
$<$Fe$>$ versus color data; whereas, the 12 Gyr SSP is a lesser fit to the
data and all the frosting models fail to match the $<$Fe$>$ versus color
data.

Another consideration is the effect of higher $\alpha$/Fe models on the
H$\beta$ diagrams.  As noted in \S3, SED models with higher ratios of the
$\alpha$ elements (e.g. Mg) will have slightly redder colors than solar
models of the same [Z/H].  Likewise, $\alpha$-enhanced models will have
lower $<$Fe$>$ indices than solar models for a constant [Z/H].  This
effectively raises the H$\beta$ value per [Fe/H] value estimated from
$<$Fe$>$.  We estimated, in section \S3, that increased [$\alpha$/Fe]
values of $+$0.3 (typical for bright ellipticals, Thomas \etal 2005), would
lead to an increase in $vz-yz$ of 0.03 (see Figure 5) and an increase in
H$\beta$ of 0.2.  This is insufficient to explain the discrepancies in
Figure 9, but is in the correct direction.  For comparison, we plot the
$\alpha$-enhanced models of Idiart \etal (2007) in Figure 9.  Again, these
models are in the correct direction to resolve the disagreement between
models and the data, but fail to predict the magnitude of the H$\beta$
index.

One important point to extract from Figure 9 is the strong correlation
between color and the H$\beta$ index.  For most spectroscopic studies (e.g.
Thomas \etal 2005), this correlation has been interpreted as reflecting
decreasing mean age with decreasing galaxy mass.  However, if we accept
old, and coeval, ages for ellipticals, than the color-H$\beta$ relation
must signal a dependence between metallicity and H$\beta$ that is stronger
than the expected increase in H$\beta$ due to an increase in the RGB
effective temperature with lower metallicity.  Again, one might be inclined
to invoke a `2nd parameter' problem for galaxy spectra where lower
metallicities involve a complex mix of blue horizontal branch stars.
However, as discussed above, this does not resolve the lack of bluer
continuum colors that such a population would express.

\section{Conclusions}

Using a narrowband color system, we have examined the impact of the
deduced younger ages for early-type galaxies in clusters proposed by
numerous spectroscopic surveys under the Lick system.  The color-magnitude
diagram is the tool of choice for testing SED model colors as one axis
(magnitude) is limited in its explanation (i.e., galaxy stellar mass)
leaving only galaxy color as open for interpretation.

\noindent We summarize our results as follows:

\begin{itemize}

\item{} The CMR for our narrowband color system is well defined and
consistent with the known changes in broadband colors.  The slope is
steepest for the $vz-yz$ `metallicity' color, that also has the lowest
internal errors and intrinsic scatter.

\item{} SED models have reached a level of sophistication that now allow
for detailed testing of integrated colors as a function of age and
metallicity.  However, as expected, a range of model parameters (in terms
of stellar population age and metallicity) can reproduce the various
color-color diagrams in our filter system.  Of course, the converse is not
true, a pair of age and metallicity values will correspond to a unique
galaxy color.  We can use this effect to test the integrity of ages and
metallicities determined by spectroscopic values by comparing predicted
model colors with a one parameter relation such as the CMR.  As these
models contain the same assumptions and calculations as the models in which
the spectroscopic groups use to convert their indices into age and
metallicity, then the results should be the same.

\item{} Converting the ages and metallicities from the Trager \etal (2008)
and Thomas \etal (2005) spectroscopic samples into colors, then comparing
them to the observed CMR, demonstrates that spectroscopically determined
ages are sharply discordant with CMR (see Figure 5).  This discrepancy is
visible in all colors (broadband and narrowband), but has the strongest
signature in our narrowband system, mostly due to the high sensitivity to
metallicity and age effects in our narrowband system and the fact that our
colors bracket the region of a galaxy's spectrum containing the H$\beta$
feature.  Most importantly, this discrepancy is resolved if model ages of
12 Gyr are assumed (see Figure 7).  Using spectroscopic ages, yet varying
metallicity, requires unrealistic super-solar global metallicity values for
galaxies.

\item{} Any possibility that inaccurate spectroscopic values are
responsible for young ages is removed when the high resolution Keck data
for Coma is considered (Trager \etal 2008).   This dataset has the highest
S/N of any spectroscopic survey, particularly for the critical H$\beta$
index, and still display the same inconsistent colors for a stellar
population of assumed young age (see Figure 8).  

\item{} On the other hand, using the same Coma data, with matching
narrowband colors, finds excellent agreement between spectral indices, such
as $<$Fe$>$ and colors.  The conflict only arises between the H$\beta$
index and colors.  There is a strong correlation between H$\beta$ and color
(bluer colors result in higher H$\beta$ values); however, none of our
experiments to add a hot stellar component to an underlying old stellar
population can raise the SED H$\beta$ values without introducing discordant
blue colors.

\end{itemize}

We are left with the conclusion that the unusual aspect to ellipticals is
not a young stellar age, but rather their high H$\beta$ values without a
significant hot component to their underlying stellar population to explain
these values.  And our numerical experiments demonstrate that if you adopt
$<$Fe$>$ values calibrated to [Fe/H] then you can reproduce the CMR with no
need for galaxy mean ages of less than 10 Gyr.  There are only two possible
interpretations for this dilemma, either 1) the H$\beta$ index is
incorrectly interpreted and the galaxies are composed of old (greater than
10 Gyr) stellar populations to match their colors and metallicities as
given by metal indices or, 2) our SED models correctly map the H$\beta$
values into age-metallicity space and the spectroscopic results are
correct, i.e., many galaxies have young stellar populations, but then the
SED models incorrectly predict all narrowband and broadband colors.

Nearly every spectroscopic study of ellipticals has revealed a significant
fraction with large H$\beta$ indices (greater than 1.6) which implies a
young mean stellar population age through interpretation by the most basic
SED models (Cervantes \& Vazdekis 2008; Caldwell, Rose \& Concannon 2003).
However, galaxies with ages less than 8 Gyr would challenge our
understanding of passive color evolution of galaxies at intermediate
redshifts (Rakos \& Schombert 1995).  While galaxy colors redden quickly
and stabilize after 8 Gyrs (Rakos, Schombert \& Odell 2008), ages less than
8 Gyrs, as seen in many spectroscopic studies, are impossible to reconcile
with the observed colors at redshifts of 0.2 to 0.4.  For example, Rakos \&
Schombert (2005) find the colors of A2218 ($z$=0.18) are consistent with a
lookback time of 2 Gyr.  However, if the mean ages from the spectroscopic
study of Coma are extrapolated to A2218, the mean colors would be 0.2 mags
bluer than those observed.

This has serious consequences to current state of our field of galaxy
evolution as every spectroscopic study of galaxies using the Lick system
has, therefore, incorrectly deduced galaxy age.  This ranges from
individual spectroscopic surveys (e.g. Smith \etal 2007; Sanchez-Blazquez
\etal 2006) to large SDSS programs (e.g. Clemens \etal 2008).  As these age
datasets are used to test our galaxy formation scenarios, a great deal of
our conclusions on the star formation history of ellipticals is in error.

{\it Note on astro-ph version}:  This version of our work released on
astro-ph deviates slightly from the version that will be published.  This
is due to, what we believe, is a repressive editorial policy that allows an
anonymous referee to replace our conclusions to match their personal views.
Thus, we have restored several sections of text to re-enforce, more strongly,
our results based on our data.  Since no one ever references our work, we
find the possible confusion in bibliography's to be moot.

\acknowledgements

Financial support from Austrian Fonds zur Foerderung der Wissenschaftlichen
Forschung and NSF grant AST-0307508 is gratefully acknowledged.  We also
acknowledge all the telescope time granted this project from NOAO (north
and south), as well as ESO.  This research has made extensive use of the
NASA/IPAC Extragalactic Database (NED) which is operated by the Jet
Propulsion Laboratory, California Institute of Technology, under contract
with the National Aeronautics and Space Administration.  Funding for the
SDSS and SDSS-II has been provided by the Alfred P. Sloan Foundation, the
Participating Institutions, the National Science Foundation, the U.S.
Department of Energy, the National Aeronautics and Space Administration,
the Japanese Monbukagakusho, the Max Planck Society, and the Higher
Education Funding Council for England.

\begin{deluxetable}{rccccccccc}
\tablecolumns{7}
\small
\tablewidth{0pt}
\tablecaption{SSP and Multi-Metallicity SED Models}
\tablehead{
\colhead{[Fe/H]} &
\colhead{$uz-vz$} &
\colhead{$vz-yz$} &
\colhead{$bz-yz$} &
\colhead{$U-B$} &
\colhead{$B-V$} &
\colhead{$V-K$} &
\colhead{$<$Fe$>$} &
\colhead{Mg$_b$} &
\colhead{H$\beta$} 
}
\startdata
\sidehead{5 Gyr SSP}
-0.7 & 0.614 & 0.289 & 0.202 & 0.201 & 0.725 & 2.387 & 1.587 & 2.174 & 2.053 \nl
-0.4 & 0.646 & 0.378 & 0.229 & 0.280 & 0.773 & 2.569 & 1.937 & 2.307 & 2.149 \nl
 0.0 & 0.712 & 0.535 & 0.276 & 0.431 & 0.855 & 2.867 & 2.458 & 2.894 & 2.080 \nl
+0.4 & 0.794 & 0.712 & 0.332 & 0.595 & 0.958 & 3.317 & 2.900 & 3.763 & 1.711 \nl
\sidehead{12 Gyr SSP}
-0.7 & 0.628 & 0.418 & 0.238 & 0.289 & 0.799 & 2.522 & 1.814 & 2.846 & 1.561 \nl
-0.4 & 0.690 & 0.540 & 0.278 & 0.415 & 0.867 & 2.813 & 2.191 & 2.933 & 1.640 \nl
 0.0 & 0.783 & 0.711 & 0.329 & 0.594 & 0.954 & 3.116 & 2.817 & 3.565 & 1.610 \nl
+0.4 & 0.909 & 0.902 & 0.381 & 0.797 & 1.054 & 3.504 & 3.219 & 4.316 & 1.272 \nl
\nl
\hline
\nl
\sidehead{5 Gyr Multi-Metallicity}
-0.7 & 0.622 & 0.263 & 0.200 & 0.185 & 0.711 & 2.363 & 1.506 & 1.995 & 2.174 \nl
-0.4 & 0.643 & 0.342 & 0.222 & 0.251 & 0.755 & 2.539 & 1.811 & 2.264 & 2.144 \nl
 0.0 & 0.686 & 0.464 & 0.260 & 0.357 & 0.823 & 2.851 & 2.231 & 2.753 & 2.046 \nl
+0.4 & 0.737 & 0.593 & 0.303 & 0.467 & 0.899 & 3.245 & 2.610 & 3.380 & 1.819 \nl
\sidehead{12 Gyr Multi-Metallcity}
-0.7 & 0.633 & 0.388 & 0.238 & 0.267 & 0.785 & 2.515 & 1.710 & 2.577 & 1.702 \nl
-0.4 & 0.671 & 0.486 & 0.266 & 0.360 & 0.839 & 2.744 & 2.059 & 2.876 & 1.660 \nl
 0.0 & 0.738 & 0.625 & 0.308 & 0.494 & 0.915 & 3.066 & 2.523 & 3.362 & 1.574 \nl
+0.4 & 0.811 & 0.762 & 0.349 & 0.624 & 0.991 & 3.413 & 2.890 & 3.918 & 1.370 \nl
\enddata
\end{deluxetable}

\end{document}